\documentclass[
 reprint,
 amsmath,amssymb,
 aps,
]{revtex4-1}

\usepackage{graphicx}
\usepackage{dcolumn}
\usepackage{bm}
\usepackage{amsmath}
\usepackage{amssymb}
\usepackage{braket}
\usepackage{physics}
\usepackage{amsthm}
\usepackage{natbib}
\usepackage{mathtools}
\usepackage[utf8]{inputenc}
\usepackage[english]{babel}
\begin{document}
\title{Approximate stabilizer rank and improved weak simulation of Clifford-dominated circuits for qudits}

\author{Yifei Huang}
\author{Peter Love}
\affiliation{Department of Physics and Astronomy, Tufts University.}
\begin{abstract}
Bravyi and Gosset recently gave classical simulation algorithms for quantum circuits dominated by Clifford operations. These algorithms scale exponentially with the number of $T$-gates in the circuit, but polynomially in the number of qubits and Clifford operations. Here we extend their algorithm to qudits of odd prime dimension. We generalize their approximate stabilizer rank method for weak simulation to qudits and obtain the scaling of the approximate stabilizer rank with the number of single-qudit magic states. We also relate the canonical form of qudit stabilizer states to Gauss sum evaluations and give an $O(n^3)$ algorithm for calculating the inner product of two $n$-qudit stabilizer states.
\end{abstract}

\maketitle

\section{\label{sec:level1}Introduction}

With the prospect of noisy intermediate scale quantum (NISQ) computers with $50-100$ qubits appearing in the next decade~\cite{boixo2018characterizing,neill2018blueprint}, determining the minimal classical cost of simulation of quantum computers has received much recent attention~\cite{bravyi2016improved,smelyanskiy2016qhipster,haner20170,boixo2017simulation,markov2018quantum}.

The Gottesman-Knill theorem shows that Clifford circuits are efficiently classically simulatable~\cite{aaronson2004improved}. Adding any non-Clifford gate creates a universal gate set~\footnote{An elegant recent presentation of this result in group-theoretic terms is given in~\cite{nebe2001invariants} and is briefly summarized in~\cite{campbell2012magic}}. One such choice for a non-Clifford gate is the $T$ gate: $T\ket{j}=e^{ij\pi/4}\ket{j},~~j\in \{0,1\}$~\cite{boykin1999universal}. Bravyi and Gosset gave a classical algorithm for simulation of quantum circuits that scales exponentially with the number of $T$-gates in the circuit but polynomially with the number of qubits and Clifford gates~\cite{bravyi2016improved}. This algorithm was further developed in~\cite{bravyi2018simulation}.

What is supplied by the addition of $T$-gates to a Clifford circuit? The fault tolerant implementation of Clifford+$T$ circuits substitutes magic states for each $T$ gate~\cite{bravyi2005universal,zhou2000methodology}. Colloquially, $T$ gates add ``magic'' to a Clifford circuit. Magic is supplied by contextuality, a longstanding source of puzzles and paradoxes in the foundations of quantum mechanics~\cite{howard2014contextuality}.

The relationship of magic to contextuality also provides a connection to quasiprobability representations of quantum mechanics~\cite{spekkens2008negativity,ferrie2008frame,ferrie2008frame}. Specifically, positivity of a quasiprobability representation is equivalent to the absence of contextuality, and such positive states, operations and measurements admit efficient classical simulation in some cases~\cite{veitch2013efficient,mari2012positive}. Classical statistical theories with an imposed uncertainty principle  can reproduce these positive quasiprobabilistic theories for Gaussian states and  qudits with $d>2$~\cite{spekkens2016quasi,bartlett2012reconstruction}. 

Pashayan {\em et al.} gave an algorithm allowing a positive quasiprobability description to include some negativity~\cite{pashayan2015estimating}. Comparing the algorithms of Bravyi and Gosset and Pashayan should shed more light on the relationship between magic, contextuality and negativity~\cite{bravyi2016improved,pashayan2015estimating}. However quasiprobability representations for qubits are distinct from their $d$-dimensional cousins~\cite{kocia2017discrete,kocia2017b,kocia2018a}. The desire to understand the relationship between magic, contextuality and negativity therefore motivates extension of the algorithm of Bravyi and Gosset to qudits with dimension greater than two. In the present paper we extend the algorithm of Bravyi and Gosset to qudits of odd prime dimension. 

The structure of the paper is as follows. In Sections~\ref{clifford} and ~\ref{BG}, we briefly introduce the necessary background. In Section~\ref{decomposition} we give the nonorthogonal decomposition of the magic state, and in Section~\ref{simulation} we give results on approximate stabilizer rank and weak simulation algorithm for qudits. We close the paper by briefly comparing our algorithm to that of ~\cite{pashayan2015estimating}. 

\section{Qudit Pauli group and Clifford gates}\label{clifford}

The Pauli and Clifford groups were first generalized beyond qubits by Gottesman~\cite{gottesman1999fault}. Assuming henceforth that $d$ is an odd prime, we define the Heisenberg-Weyl operators:
\begin{equation}
D_{\vec x}=\tau^{xz}X^x Z^z,
\end{equation}
where $X\ket{j}=\ket{j\oplus1}$, where $\oplus$ denotes addition modulo $d$, $Z\ket{j}=\omega^j\ket{j}$, $\vec x=(x,z)$, where $x$ and $z$ are integers modulo $d$, $\omega=\exp(2\pi i/d)$ and $\tau=e^{(d+1)\pi i/d}=\omega^{2^{-1}}$.
The Heisenberg-Weyl operators form a group whose product rule follows from the Heisenberg-Weyl commutation relation $\omega XZ= ZX$:   
\begin{equation}
D_{\vec x_1}D_{\vec x_2} =\tau^{\langle\vec x_1\cdot\vec x_2\rangle}  D_{\vec x_1+\vec x_2}
\end{equation}
where $\langle\vec x_1\cdot\vec x_2\rangle$ is the symplectic inner product:
$\langle\vec x_1\cdot\vec x_2\rangle = z_1 x_2 - x_1 z_2$.

The generators of the Clifford group on qudits are $P$, $H$ and $CNOT$, where $P\ket{j}=\omega^{j(j-1)/2}\ket{j}$, $H\ket{j} =d^{-1/2}\sum_k \omega^{jk}\ket{k}$ and $CNOT\ket{j,k}=\ket{j,k\oplus j}$. We can also write any single qudit Clifford unitary as $C_{F,\vec\chi}=D_{\vec\chi}U_F$, where $\vec\chi=(x,z)$ and $F$ is a $2\times2$ matrix with entries modulo $d$. We will make particular use of matrices $C_{\gamma,\vec\chi}=D_{\vec\chi}U_\gamma$ for $U_\gamma\ket{k}=\tau^{\gamma k^2}\ket{k}$. The order of $C_{\gamma,\vec\chi}$ is $d$. The Clifford group is reviewed in more detail in Appendix~\ref{appcliff}.

Qudit stabilizer states can be prepared from a logical basis state by a qudit Clifford circuit. The Gottesmann-Knill theorem generalizes to qudits and qudit stabilizer computations allow efficient classical simulation~\cite{gottesman1999fault}. Qudit stabilizer states possess canonical forms in the logical basis just as in the qubit case~\cite{nest2008classical,dehaene2003clifford,hostens2005stabilizer}. 

The remaining generalization we require is an efficient classical algorithm for obtaining the inner product of two stabilizer states. This is required by the algorithm of Bravyi and Gosset and the qubit case was given in~\cite{bravyi2016improved}. We give an $O(n^3)$ algorithm for the inner product of two $n$-qudit stabilizer states based on Gauss sums in Appendix~\ref{stabilizer}. 

The qudit $T$-gate was defined in~\cite{howard2012qudit, campbell2012magic} as a diagonal gate $U_T$ that maps Pauli operators to Clifford operators. Its action is specified by the image of $X=D_{(1,0)}$ under $U_T$.  Magic states are then eigenvectors of this image. Let the eigenstate of $X$ with eigenvalue $\omega^k$ be $\ket{+_k}$, then the magic states are $U_T\ket{+_k}$. This approach is that taken by Howard in~\cite{howard2012qudit}.

The image of $X$ under $U_T$ can be written (up to a phase) as $C=XP^\gamma Z^\xi$ for $\gamma$, $\xi$ integers modulo $d$. The effect of nonzero $\xi$ is simply to reorder the eigenvectors and hence we can choose $\xi=0$. Similarly, the eigenvectors for $\gamma>1$ and $\gamma=1$ are related by application of $P^{\gamma-1}$, a Clifford operator. We can therefore specialize to the case $\gamma=1$ and $\xi=0$, and the gate with action:
\begin{equation}
C_d=M_dXM_d^{\dagger}= 
\begin{cases}
e^{2\pi i/9}XP. & d=3.\\
\omega^{-\bar{3}}XP. & d>3. \label{C}
\end{cases}
\end{equation}
where $\bar{3}$ indicates the multiplicative inverse of $3$ modulo $d$. This is the gate defined by Campbell {\em et al.} in~\cite{campbell2012magic}. The qudit magic states are reviewed in more detail in Appendix~\ref{magic1}.

The definition of magic states allows one to replace a Clifford+$T$ circuit with a Clifford circuit with injected magic states~\cite{,zhou2000methodology,bravyi2005universal}. This construction was extended to qudits in~\cite{howard2012qudit} and we review it in Appendix~\ref{gadget}. In Section~\ref{BG} we will review the Bravyi-Gosset algorithm for qubits which we will generalize to qudits.

\section{The Bravyi-Gosset Algorithm\label{BG}}

Bravyi and Gosset gave algorithms for both weak and strong simulation in~\cite{bravyi2016improved,bravyi2018simulation}. A {\em strong simulation} outputs the probability of measuring output $x$ from a given Clifford+$T$ circuit. A {\em weak simulation} algorithm generates samples from the probability distribution over outputs of a given Clifford+$T$ circuit. Here we review the weak simulation algorithm. A brief summary of relevant features of the strong simulation algorithm is given in Appendix~\ref{AppBG}.

The key advantage of weak simulation is that one can sample from a $\tilde{P}_{out}(x)$ that is close enough to the actual $P_{out}(x)$. Bravyi and Gosset devised a method to approximate the $t$-qubit magic state $\ket{{A^{\otimes t}}}$, where $\ket{A}=2^{-1/2}(\ket{0}+e^{i\pi/4}\ket{1}$, with a superposition of $<2^t$ stabilizer states. 

 The {\em approximate stabilizer rank} $\chi'$ is defined as the minimal stabilizer rank (defined in~\cite{bravyi2016trading} and reviewed in Appendix~\ref{AppBG}) of a state $\ket{\psi}$ that satisfies $\abs{\braket{\psi}{A^{\otimes t}}}\geq 1-\delta$. A close approximation to the tensor product of magic states means a close approximation to the action of a Clifford+$T$ circuit realized by magic state injection~\cite{bravyi2016improved}. Therefore, $\tilde{P}_{out}(x)$ will be close enough to $P_{out}(x)$ if $\delta$ is small enough.

The sampling procedure given by Bravyi and Gosset relies on standard computations of stabilizers. The extension of such computations to $d>2$ have long been well understood~\cite{gottesman1999fault}. We will therefore refer the reader to~\cite{bravyi2016improved} for details of these procedures which, {\em mutatis mutandis}, can be applied in the qudit case, and focus on the approximate stabilizer rank.

We begin by reviewing the approximate stabilizer rank construction from~\cite{bravyi2016improved}. From the magic state $\ket{A}$ defined above one can construct the equivalent magic state:
\begin{equation}\label{HMagic}
\ket{H}=e^{-\pi i/8}PH\ket{A}=\cos(\pi/8)\ket{0}+\sin(\pi/8)\ket{1}.
\end{equation}
The state $\ket{H}$ can be decomposed into a sum of non-orthogonal stabilizer states as follows:
\begin{equation}\label{Hnonorth}
\ket{H}=\frac{1}{2\cos{\pi/8}}(\ket{\tilde{0}}+\ket{\tilde{1}}) 
\end{equation}
where $\ket{\tilde{0}}=\ket{0}$ and $\ket{\tilde{1}}=\frac{1}{\sqrt{2}}(\ket{0}+\ket{1})$. Then $\ket{H^{\otimes t}}$ can be rewritten as
\begin{equation}
\ket{H^{\otimes t}}=\frac{1}{(2\cos(\pi/8))^t}\sum_{x\in \mathcal{F}_2^t} \ket{\tilde{x}}    
\end{equation}

The weak simulation algorithm reduces the number of stabilizer states required by approximating $\ket{H^{\otimes t}}$. This approximation $\ket{H^{\otimes t*}}$ is constructed by taking a subspace $\mathcal{L}$  of $\mathcal{F}_2^t$:
\begin{equation}
\ket{H^{\otimes t*}}=\frac{1}{(2\cos(\pi/8))^t}\sum_{x\in \mathcal{L}} \ket{\tilde{x}}     
\end{equation}
The stabilizer rank of this approximation state is the number of elements in $\mathcal{L}$, which is $2^k$. The random subspace $\mathcal{L}$ is chosen so that $\ket{H^{\otimes t*}}$ satisfies:
\begin{equation}
\braket{H^{\otimes t*}}{H^{\otimes t}}\leq 1-\delta.  \end{equation}

It is useful to discuss the subspaces of $\mathcal{F}_2^t$ in the language of $d$-ary linear codes. $\mathcal{L}$ is a $k$-dimensional binary linear code which can be specified by $k$ generators of length $t$. These generators can be written in a standard form as a $k\times t$ matrix $\{1_k|G\}$ where $1_k$ is the $k\times k$ identity matrix and $G$ is a $k\times (t-k)$ matrix. Sampling random subspaces of $\mathcal{F}_2^t$ is therefore equivalent to sampling matrices $G$.

The algorithm of Bravyi and Gosset achieves an improved scaling of $\cos(\pi/8)^{-2t}\simeq2^{0.23t}$ for weak simulation over $2^{0.47t}$ for strong simulation. In section \ref{decomposition} and \ref{simulation}, we will see more details of how to bound the scaling while we extend this approximate rank and weak simulation scheme to qudits.

\section{Nonorthogonal decompositions of qudit Magic states} \label{decomposition}

The qudit magic state we want to decompose is an eigenvalue one eigenstate of the Clifford operator $C_d$ as defined by eq.~(\ref{C}). We choose a stabilizer state $\ket{\tilde 0}$ with non-zero inner product with the magic state and act on it with powers of $C_d$ to obtain $d$ stabilizer states $\{\ket{\tilde j}=C_d^{j}\ket{\tilde {0}},j=0,...,d-1\}$. We know these stabilizer states are distinct because if any pair were equal then the original state $\ket{\tilde {0}}$ would be an eigenstate of the Clifford operator and hence a magic state. The sum of these $d$ states form a decomposition of the magic state (up to a possible global phase). Because $C_d$ has order $d$ this state is by construction an eigenvalue one eigenstate of $C_d$.

The $d$ stabilizer states in the decomposition form an {\em orbit} around the magic state. This construction was discussed previously in~\cite{howard2015maximum}. There are $d(d+1)$ single-qudit stabilizer states~\cite{wootters1987wigner}, partitioned into $d+1$ orbits, each orbit giving a decomposition of the magic state. Every state in each orbit has the same overlap with the magic state: 
\begin{equation}
\braket{\tilde{j}}{M_d} =\bra{\tilde{j}}C_d^\dagger C_d\ket{M_d}= \braket{\widetilde{j+1}}{M_d} \label{property}
\end{equation}
where the qudit magic state is $\ket{M_d}=M_d\ket{+}$. This property is a generalization of  $\braket{\tilde{0}}{H}=\braket{\tilde{1}}{H}=\cos{\pi/8}$ for the qubit case. The overlaps of the elements of the nonorthogonal basis are given by: $\abs{\braket{\tilde{0}}{\tilde{j}}}=\frac{1}{\sqrt{d}}$ for all $j$s, i.e.:
\begin{equation}\label{noboverlap}
\abs{\braket{\tilde{j}}{\tilde{k}}}^2=\frac{1+(d-1)\delta_{j,k}}{d}.
\end{equation} 
This expression is that for states in a SIC-POVM, and the construction here is similar to the generation of such states from a fiducial state~\cite{fuchs2017sic,zhu2010sic}. Here we only obtain $d$ states, however. See Appendix~\ref{Zeval} for the evaluation of the phase of $\braket{\tilde{j}}{\tilde{k}}$. 

The states $\ket{+_p}=Z^p\ket{+}$ are representatives of the $d$ orbits, each of which generated by $C_d$. This is because $C^a_d\ket{+_p}\neq\ket{+_q}$ for any $a,p,q$, which follows simply from the action of $C_d$ in the logical basis. $C_d$ applies phases quadratic in $j$ to $\ket{j}$ followed by a shift. This cannot be equal to a state generated from $\ket{+}$ by any power of $Z$, which can only apply phases linear in $j$ to $\ket{j}$. 

From the orbit representatives we can determine the inner product of the states in the orbit with the magic state. This is given by:
\begin{equation}
\alpha = \bra{+}Z^{-p}\ket{M_d} = \bra{+}Z^{-p}M_d\ket{+}=\frac{1}{d}{\rm Tr}(Z^{-p}M_d).
\end{equation}
This is a cubic gauss sum which can be written:
\begin{equation}\label{inprod}
\alpha=\frac{\omega^{\frac{1}{d}{d \choose 4}-p}}{d}\sum_{l=0}^{d-1}\omega^{\bar 6l(l^2+\psi(p,d))} ~~~ d>3. 
\end{equation}
For the $d=3$ case, the magnitude and phase of this cubic Gauss sum, and $\phi(p,d)$, are computed in Appendix~\ref{appinner}. The sum is real, although not necessarily positive. Although we  do not obtain a closed form for this sum, we can compute the integer value of $p$ which maximizes its absolute value for a given $d$. These values are tabulated for small $d$ in Table~\ref{sumtab}.

The complete form of the nonorthogonal decomposition is:
\begin{equation}
\ket{M_d}=\pm\frac{\omega^{\frac{1}{d}{d \choose 4}-p}}{d |\alpha|}\sum_{j}C_d^{j}\ket{\tilde 0}.   \label{N}
\end{equation}
which is the generalization of eq.~(\ref{Hnonorth}) to arbitrary $d$. 

\section{Weak Simulation and Approximate Stabilizer Rank}\label{simulation}

In order to get an approximation for $\ket{M^{\otimes t}}$, we can follow the method of Bravyi and Gosset for the qubit case, taking a $k$-dimensional subspace of $\mathcal{F}_d^t$:
\begin{equation}
\ket{M^{\otimes t*}}=\ket{\mathcal{L}}=\frac{1}{\sqrt{d^kZ(\mathcal{L})}}\sum_{x\in\mathcal{L}}\ket{\tilde{\vec{x}}}   
\end{equation}
Here we label the state by $\mathcal{L}\subset \mathcal{F}_d^t$, a $k$ dimensional code subspace of $\mathcal{F}_d^t$ and $Z(\mathcal{L})$ is a normalization factor. Comparison with eq. (\ref{N}) shows that $Z(F_d)=d|\alpha|^2$. We require:
\begin{equation}
\abs{\braket{\mathcal{L}}{M^{\otimes t}}}^2=\frac{d^{k}\abs{\alpha}^{2t}}{Z(\mathcal{L})} \geq 1-\delta  \label{condition} 
\end{equation}
for a given $\delta$, where the first equality follows from eq. (\ref{property}) and where:
\begin{equation}
\begin{split}
Z(\mathcal{L}) &=\sum_{\vec{x}\in \mathcal{L}}\bra{\tilde{0}^{\otimes{t}}}C_{\vec{x}}\ket{\tilde{0}^{\otimes{t}}}   \label{Z(L)}
\end{split}
\end{equation}
for $C_{\vec{x}}=C^{x_1}\otimes C^{x_2}...\otimes C^{x_t},~x_i\in \mathcal{F}_d$. 

Selection of the subspace $\mathcal{L}$ depends on two factors. First, we choose the dimension of $\mathcal{L}$ by setting $k$:
\begin{equation}\label{k}
k = \lceil 1-2t\log_d |\alpha| -\log_d \delta\rceil.    
\end{equation}
Note that the maximum precision that can be required from the method for given $t$ is obtained by setting $k=t$, so that $\delta_{\rm max} = 2^{-t(1+2\log_d|\alpha|)+1}$.

Next we find an $\mathcal{L}$ for which $Z(\mathcal{L})$ is not too large. The probability of obtaining a small enough $Z(\mathcal{L})$ can be  analyzed as in~\cite{bravyi2016improved} by evaluating the expectation value of $Z(\mathcal{L})$ over all possible $\mathcal{L}\in \mathcal{F}_d^t$:
\begin{equation}
\begin{split}
E(Z(\mathcal{L}))&=1+\sum_{\vec{x}\in \mathcal{F}_d^t/\{0\}} \bra{\vec{\tilde{0}}^t}C_{\vec{x}}\ket{\vec{\tilde{0}}^t}E(I_{\mathcal{L}}(\vec{x}))\\
&=1+\frac{(d^k-1)}{(d^t-1)}(Z(F_d)^t-1)\\
&=(1+\frac{d^k-1}{d^t-1}(d^t\abs{\alpha}^{2t}-1))
\\
&\leq (1+d^k\abs{\alpha}^{2t}) \label{expvalue}
\end{split}
\end{equation}
Here $I_{\mathcal{L}}(\vec{x})$ is a indicator function, {\em i.e.}, it is equal to 1 when $x\in\mathcal{L}$ and 0 otherwise. The second equal sign stands because the expectation value of $I_{\mathcal{L}}(x)$ for a fixed $x$ is $\frac{d^k-1}{d^t-1}$ and 
\begin{equation}
\begin{split}
\sum_{x\in \mathcal{F}_d^t/\{0\}} \bra{\vec{\tilde{0}}^t}C_{\vec{x}}\ket{\vec{\tilde{0}}^t}&=\sum_{x\in \mathcal{F}_d^t} \bra{\vec{\tilde{0}}}C_{\vec{x}}\ket{\vec{\tilde{0}}}-1 \\
&=\left(\bra{\tilde{0}}\sum_{x=0}^{d-1}C^x\ket{\tilde{0}}\right)^t-1. 
\end{split}
\end{equation}

From eq. (\ref{k}) we have $d^k|\alpha|^{2t}=O(1)$ so $E(Z(\mathcal{L}))=O(1)$. Therefore from Markov's inequality we obtain
\begin{equation}
\begin{split}
&{\rm Prob}\biggl[Z(\mathcal{L})\leq (1+d^k\abs{\alpha}^{2t})(1+\delta)\biggr]\\
&> 1-\frac{E(Z(\mathcal{L}))}{(1+d^k\abs{\alpha}^{2t})(1+\delta)}
\geq 1-\frac{1}{1+\delta}>\delta.
\end{split}
\end{equation}
Randomly choosing $\delta^{-1}$ subspaces gives an $\mathcal{L}$ such that:
\begin{equation}
Z(\mathcal{L})\leq (1+d^k\abs{\alpha}^{2t})(1+\delta)    
\end{equation}
and hence satisfying eq.(\ref{condition}), with high probability.

The upper bound for the approximate stabilizer rank of a $t$-qudit magic state given by the above method is:
\begin{equation}
\chi'(t)=d^k=O(\delta^{-1}\abs{\alpha}^{-2t}).
\end{equation}
In the qubit case an explicit sum formula was given for $Z(\mathcal{L})$ with $2^k$ terms, and hence the cost of evaluating $Z(\mathcal{L})$ is $O(2^k)$. What is the cost of evaluating $Z(\mathcal{L})$ for arbitrary $d$? In Appendix~\ref{Zeval} we give an explicit formula for $Z(\mathcal{L})$ as a sum of products, and hence the cost of evaluating $Z(\mathcal{L})$ for arbitrary $d$ is $O(d^{k+1})$.

\begin{table}[h!]
    \centering
    \begin{tabular}{c|c|c|c|c|c}
        $d$ &  $M_d$ & $p$ & $\abs{\alpha(d)}$ &$\abs{\alpha(d)}$ & $d^{\kappa t}$\\
        \hline
        2 & ${\rm diag}(1,e^{i\pi/4})$ & $0$ & $\cos\pi/8$&$0.92388$&$2^{0.23t}$ \\
        3 & ${\rm diag}(e^{2\pi i/9},1,e^{-2\pi i/9})$ & $0$ & $\frac{1+2\cos(2\pi/9)}{3}$ &$0.84403$&$3^{0.32t}$\\
        5 & ${\rm diag}(\omega^{-2},\omega,\omega^{-1},\omega^{-2},\omega^{-1})$ & $4$ &$\frac{3+2\cos(2\pi/5)}{5}$ &$0.723607$&$5^{0.41t}$\\
        7 & ${\rm diag}(\omega^{3},\omega^{-2},1,\omega^{3},\omega^{1},\omega^{2},1)$ & $3$ &$\frac{1+6\cos(2\pi/7)}{7}$ &$0.677277$& $7^{0.40t}$
    \end{tabular}
    \caption{The matrices $M_d$, optimal value of $p$ and approximate stabilizer rank scaling comparison for $d=2,3,5,7$. Here $\kappa=-2\log_d\alpha$ so that $d^{\kappa t}=\alpha^{-2t}$. Here the $\omega$ for $d=5$ and $d=7$ rows are $e^{2\pi i/5}$ and $e^{2\pi i/7}$ respectively.\label{sumtab}}
\end{table}

\section{Discussion}

The motivation to study the qudit generalizations of stabilizer rank algorithms such as those in~\cite{bravyi2016improved,bravyi2018simulation} is to enable comparison with other simulation algorithms. In~\cite{pashayan2015estimating}, the authors apply Monte Carlo sampling on trajectories of the quasiprobability representation to estimate the probability of a measurement outcome. They find the hardness of this strong simulation depends on the total negativity (Negativity of the inputs, gates and measurements) of the circuit. Specifically the cost of the algorithm scales with the square of the total negativity. 

For Clifford+$T$ circuits that are gadgetized so that the circuit is realized by Clifford gates with magic state injection, the negativity of the circuit only comes from the ancilla inputs of magic states. If we apply the method of~\cite{pashayan2015estimating} to the gadgetized circuit with an input of $t$-qutrit magic states, the cost scales as $3^{0.84t}$. This result is obtained by calculating the negativity of a single-qutrit magic state. 

In the present paper, we obtain a scaling of $3^{0.32t} $ for weak simulation of qutrit Clifford+$T$ circuits. This shows that weak simulation using the approximate rank method has superior scaling to strong simulation using the method of~\cite{pashayan2015estimating}. A stabilizer rank based strong simulation algorithm for qudits would require new results on exact stabilizer rank of qudit magic states, a topic for future work. Recent progress in extending the qubit case has been reported in~\cite{bravyi2018simulation}, and improvements to Pashayan's algorithm using a discrete systems generalization of the stationary phase approximation were given in~\cite{kocia2018b}.

It should be noted that one should not think of weak simulation as easy and strong simulation as hard. The difficulty of weak and strong simulation is a property of the distribution being sampled or computed. In some cases, such as quantum supremacy, we expect the difficulty of weak and strong simulation to coincide~\cite{boixo2017simulation}.

If we consider negativity and stabilizer rank as two measures of quantumness, we can see that they differ. Bravyi {\em et al.}~\cite{bravyi2016trading} conjectured that the magic state has the smallest stabilizer rank out of the non-stabilizer states. However, the quasi-probability of the magic state has the largest negativity. In fact, Howard and Campbell also noticed this disagreement between stabilizer rank and robustness of magic~\cite{howard2017application}. It is worth noting the differences between stabilizer rank and approximate stabilizer rank. Namely, the approximate stabilizer rank seems to agree with other measures of quantumness such as negativity or robustness of magic in that it reaches a maxima at the magic state and a minima on stabilizer states. The exact stabilizer rank does not share these properties. This makes the investigation of the difference between exact and approximate stabilizer rank interesting.

\section*{Acknowledgements}
The authors thank Robert Lemke-Oliver, Dmitris Koukoulopoulos, Juspreet Sandhu, Elizabeth Crosson, Stephen Jordan and David Gosset for helpful discussions. This work was supported by  NSF award number PHY 1720395 and from Google Inc.

\appendix

\section{The Qudit Clifford Group\label{appcliff}}

We recall that $d$ is an odd prime. In a $d$ dimensional system the Pauli operators $X$ and $Z$ are defined as:
\begin{equation}
X=\sum_{j\in F_d} \ket{j\oplus 1}\bra{j} \qquad Z=\sum_{j\in F_d} \omega^{j} \ket{j}\bra{j},
\end{equation}  
where $\omega=\exp(2\pi i/d)$. These operators obey the Heisenberg-Weyl commutation relation:
\begin{equation}
\omega XZ= ZX.   
\end{equation}

In $d$ dimensions the Weyl-Heisenberg displacement operators are defined by:
\begin{equation}
D_{\vec x}=\tau^{xz}X^x Z^z,
\end{equation}
where $\vec x=(x,z)$,$\tau=e^{(d+1)\pi i/d}=\omega^{2^{-1}}$. The qubit Pauli operators are recovered from this expression for $d=2$, with $D_{(1,0)}=X$, $D_{(0,1)}=Z$ and $D_{(1,1)}=-Y$. The Heisenberg-Weyl operators form a group with multiplication rule:
\begin{equation}
D_{\vec x_1}D_{\vec x_2} =\tau^{\langle\vec x_1\cdot\vec x_2\rangle}  D_{\vec x_1+\vec x_2}
\end{equation}
where $\langle\vec x_1\cdot\vec x_2\rangle$ is the symplectic inner product:
\begin{equation}
\langle\vec x_1\cdot\vec x_2\rangle = z_1 x_2 - x_1 z_2
\end{equation}
For $d>2$ the Weyl-Heisenberg operators are unitary but not generally Hermitian. 
  
In the qubit case, the Clifford gates map Pauli operators to Pauli operators. In the qudit case Clifford gates map Weyl-Heisenberg operators to one another. The generators of the Clifford group are defined so that the Hadamard gate maps $X\rightarrow Z$ and the phase gate maps $X\rightarrow XZ$. The generators of the single-qubit Clifford group are:
\begin{equation}
H=\frac{1}{\sqrt{2}}\begin{pmatrix}
1&1\\1&-1
\end{pmatrix},~~~~~~
P=\begin{pmatrix}
1&0\\0&i
\end{pmatrix}.
\end{equation}
The $d$-dimensional Clifford operators are generated by:
\begin{equation}
P=\sum_{j\in F_d} \omega^{j(j-1)/2} \ket{j}\bra{j} \qquad H=\sum_{j,k}\omega^{jk}\ket{j}\bra{k}/\sqrt{d},
\end{equation}
and:
\begin{equation}
CNOT=\sum_{j} \ket{j}\bra{j}\otimes X^{j}.
\end{equation}

The single-qudit Clifford group is isomorphic to the semidirect product group of $SL(2,Z_d)$~\footnote{$SL(2, Z_d)$ is the group of $2\times 2$ matrices with entries from $Z_d$ and determinant $1$.} and $(Z_d)^2$~\cite{appleby2009properties,zhu2010sic}. 

We can represent the Clifford group using a $2\times 2$ matrix $F$ and a $2$ vector $\vec \chi$, both with entries in $Z_d$:
\begin{equation}
\mathcal{C}=\left\{C_{(F|\vec\chi)}|F\in SL(2,Z_d), \vec\chi\in{Z_d}^2\right\}
\end{equation}

Specifically, a Clifford unitary is given as follows:
\begin{equation}
C_{(F|\vec\chi)}=D_{\vec\chi}U_F, \label{Clifford Group1}
\end{equation}
Where if: 
\begin{equation}
F=\begin{bmatrix}\alpha & \beta\\ \gamma & \delta \end{bmatrix},~~~~~\vec\chi=\begin{bmatrix} x \\ z \end{bmatrix},
\end{equation}
then:
\begin{equation}
U_F=  \frac{1}{\sqrt{d}}\sum_{j,k=0}^{d-1} \tau^{\beta^{-1}(\alpha k^2-2jk+\delta j^2)}\ket{j}\bra{k},     \label{Clifford Group2}
\end{equation}
if $\beta\neq 0$ and
\begin{equation}
U_F=  \sum_{k=0}^{d-1} \tau^{\alpha \gamma k^2}\ket{\alpha k}\bra{k}.    
\label{Clifford Group20}
\end{equation}
if $\beta=0$~\cite{zhu2010sic}.

The multiplication rule is:
\begin{equation}
C_{(F_1|\vec\chi_1)}C_{(F_2|\vec\chi_2)}=\tau^{\langle\vec\chi_1\cdot F\vec\chi_2\rangle} C_{(F_1F_2|\vec\chi_1+F_1\vec\chi_2)}.
\end{equation}
The action of the Clifford operators on the Heisenberg-Weyl operators in this representation can be given as follows:
\begin{equation}
C_{(F|\vec\chi)}D_{\vec{x}}C_{(F|\vec\chi)}^{\dagger}=\omega^{\vec\chi\cdot\vec{x}}D_{F\vec x} 
\end{equation}

In particular we are interested in Clifford operations defined by matrices of the form:
\begin{equation}
F_\gamma=\begin{bmatrix}1 & 0 \\ \gamma & 1 \end{bmatrix}
\end{equation}
and we introduce the notation:
\begin{equation}
C_{\gamma,\vec\chi}=C_{\begin{bmatrix}1 & 0 \\ \gamma & 1 \end{bmatrix},\begin{bmatrix}x\\z \end{bmatrix}}
\end{equation}
for $\vec\chi=(x,z)^T$. From Table I in Zhu~\cite{zhu2010sic} the order of any element $C_{\gamma,\vec\chi}$ is $d$. Clearly $X$, $P$ and $Z$ are order $d$. For $d=2$ $H$ is order $2$ and for $d>2$ $H$ is order $4$. 

The generators $H$ and $P$ are given by:
\begin{equation}
F_H=\begin{pmatrix}0&d-1\\1&0\end{pmatrix},~~\vec\chi_H=(0,0)^T  
\end{equation}
which follows from $HXH^\dagger=Z$ and $HZH^\dagger=X^{-1}$ and:
\begin{equation}
F_P=\begin{pmatrix}1&0\\1&1\end{pmatrix},~~\vec\chi_P=(0,(d-1)/2,)^T.
\end{equation}
These expressions for $H$ and $P$ allow us to construct the $F$ and $\vec\chi$ for any single qudit Clifford operation expressed as a word on the generators $H$ and $P$.

\section{Qudit Magic states and $T$ gates}\label{magic1}

To go beyond Clifford group computation it is useful to introduce the Clifford hierarchy, which classifies unitary operators by their action on the Pauli group. The Clifford hierarchy was defined by Gottesman and Chuang in \cite{gottesman1999demonstrating}:
\begin{equation}
 \mathcal{C}(k+1)=\big\{U|UPU\in\mathcal{C}(k), P\in\mathcal{P} \big\}~~(k\geq 0).
\end{equation}
The first level of the Clifford hierarchy is the Pauli group $\mathcal{C}(1)=\mathcal{P}$. The Clifford group is the second level of the hierarchy, unitary operators that map the Pauli group to itself. Note that elements of the Pauli group are themselves elements of the first level of the Clifford hierarchy. The third level of the Clifford hierarchy are operators that map Pauli operators to Clifford operators. The qubit $T$ gate is such an operator because $TXT^\dagger =PHP^2H$, a non-Pauli element of the second level of the Clifford hierarchy.

Bravyi and Kitaev first proposed qubit magic states in \cite{bravyi2005universal}. They define magic states as the image of $\ket{H}$ and $\ket{T}$ under single-qubit Clifford gates, where $\ket{H}$ is defined by eqn.~\ref{HMagic} and $\ket{T}$ by
\begin{equation}
\ket{T}=\cos\beta\ket{0}+\sin\beta e^{i\pi/4}\ket{1}, 
\end{equation}
for $\cos(2\beta)=\frac{1}{\sqrt{3}}$. $\ket{H}$ is the eigenstate of the Hadamard gate $H$ and $\ket{T}$ is the eigenstate of the product of Hadamard and Phase gate $PH$.

Any magic state is equivalent as a resource to any other state obtainable from it by a Clifford operation. We can define magic states more generally as the eigenstates of Clifford operations and obtain them as follows. Taking any $H$-type magic state $\ket{H}$, we have
\begin{equation}
UHU^{\dagger}U\ket{H}=UH\ket{H}=\lambda U\ket{H}
\end{equation}
where $\lambda$ is the eigenvalue of $H$ and $U$ is a Clifford gate. This means that $U\ket{H}$ is the eigenstate of a new Clifford operator $UHU^{\dagger}$. The same is true for $T$-type magic states.

Campbell {\em et al.}~\cite{campbell2012magic} used this relationship between magic states and eigenvectors of Clifford operators to extend the definition of magic states to qudits~\cite{campbell2012magic}. Concurrently, equivalent extensions were obtained by Howard and Vala~\cite{howard2012qudit}. 

\subsection{Qudit $T$ gates}

Campbell {\em et al.}~\cite{campbell2012magic} define sets of gates $\mathcal{M}_d^m$ containing all gates $M$ with the following properties:
\begin{enumerate}
    \item $M$ is diagonal
    \item $M^{d^m}=1$
    \item $\det M=1$ so that $M\in SU(d)$
    \item $M$ is in the third but not the second level of the Clifford hierarchy.
\end{enumerate}
Amongst this set of gates is the canonical $\mathcal{M}_d$ gate
\begin{equation}
M_d=\sum_{j} \text{exp}(i2\lambda_{j}\pi/d^{m})\ket{j}\bra{j} \label{M state'}
\end{equation}
Which is defined so that it maps the $X$ operator to a Clifford operator proportional to $XP$:
\begin{equation}
C_d=M_dXM_d^{\dagger}= 
\begin{cases}
e^{2\pi i/9}XP & d=3,\\
\omega^{-\bar{3}}XP & d>3. \label{C'}
\end{cases}
\end{equation}
Here $\bar{3}$ is the multiplicative inverse of $3$ modulo $d$. This Clifford operator has order $d$.

This condition, and the condition $\det M=1$, gives the following form for the $\lambda_j$ (See Appendix A of~\cite{campbell2012magic}):
\begin{equation}
\lambda_{j}=d^{m-2}\left[d\binom{j}{3}-j\binom{d}{3}+\binom{d+1}{4}\right]. \label{lambda'}
\end{equation}
The parameter $m$ determines the order $d^m$ of the operator $M$. For $d=3$ the form above is valid when $m\geq 2$. For $d>3$ it is valid when $m\geq 1$. 

By definition $M$ maps $X$, a generalized Pauli operator, to a non-Pauli Clifford operator and so is in the third, but not the second, level of the Clifford hierarchy. We can therefore think of $M$ as a generalized $T$ gate. 

From the definition of the matrix $M$ in (\ref{M state'}), we have for $d=3$ and $m=2$: 
 \begin{equation}
M_3={\rm diag}\begin{pmatrix}e^{i2\pi /9}, & 1, &  e^{-i2\pi /9}\end{pmatrix},
 \end{equation}
 and
 \begin{equation}
M_5={\rm diag}\begin{pmatrix}e^{-\frac{4\pi i}{5}},  & e^{\frac{2\pi i}{5}},  &  e^{-\frac{2\pi i}{5}},  & e^{-\frac{4\pi i}{5}},  & e^{-\frac{2\pi i}{5}} \end{pmatrix},
\end{equation}
 for $d=5$ and $m=1$ where $\omega=e^{2\pi i/5}$. The qudit version of the $T$ gate $M$, is further generalized in \cite{howard2012qudit}, which we will discuss below. 

The $T$ gate is also sometimes called the $\pi/8$ gate because 
\begin{equation}
T=e^{-i\pi/8}\begin{pmatrix} e^{i\pi/8} & 0 \\ 0 & e^{-i\pi/8}\end{pmatrix}.
\end{equation}
Vala and Howard developed the qudit versions of this gate concurrently with Campbell et al's development of qudit magic states~\cite{campbell2012magic,howard2012qudit}. The results are equivalent and we give the details of the relationship between them here.

Vala and Howard parameterize the set of diagonal gates on a single qudit as follows:
\begin{equation}
U_{v}=U(v_0,v_1,...,v_{p-1})=\sum_{j=0}^{d-1} \omega^{v_k}\ket{k}\bra{k} (v_k\in Z_{d}). \label{U_v}
\end{equation}
All diagonal gates fix $D_{(0,1)}$ and so their action is completely determined by $U_v D_{(1,0)} U_v^\dagger=U_v X U_v^\dagger$. This parallels the development of Campbell et al. who considered the action of their canonical gate $M$ on the operator $X$ and insisted that the result of that action was $\propto XP$.

Vala and Howard proceed more generally, computing the action of these diagonal matrices:
\begin{equation}
U_{v}D_{(x|z)}U_{v}^{\dagger}=D_{(x|z)}\sum_k\omega^{(v_{k+1}-v_k)}\ket{k}\bra{k}. \label{UDU}
\end{equation} 
Given $U_v$ is diagonal, only $U_vD_{(1|0)}{U_v}^{\dagger}$ is nontrivial. 

Vala and Howard then consider the case that $U_v$ is in the third level of the Clifford hierarchy so that the image of $X$ can be written (c.f. eq (18) in~\cite{howard2012qudit}):
\begin{equation}
U_vX{U_v}^{\dagger}=\omega^{\epsilon'}C_{\gamma',(1,z')^T}. \label{UDU2}
\end{equation}
where $\epsilon', \gamma', z'\in Z_d$. The right hand side here is the most general form allowed because eqn.~(\ref{UDU}) implies that the image of $X$ must be $X$ times a diagonal Clifford operator, and the most general form of a diagonal Clifford operator has $\vec\chi = (0,1)$ and $\beta=0$, $\alpha=1$. Combining equation (\ref{UDU}) and (\ref{UDU2}), one obtains (c.f. eq. (19) in~\cite{howard2012qudit}):
\begin{equation}
X\sum_k\omega^{(v_{k+x}-v_k)}\ket{k}\bra{k}=\omega^{\epsilon'}C_{\gamma,(1,z')^T}.
\end{equation}

Vala and Howard then solve for $U_v$ with these $3$ parameters.
\begin{equation}
v_k=\bar{12}k\{\gamma'+k[6z'+(2k-3)\gamma']\}+k\epsilon', \label{generalform1}
\end{equation}
This analysis is equivalent to that performed in Campbell et al. \cite{campbell2012magic}, Appendix A. 

The $d=3$ case as usual presents some special difficulties. In the Campbell analysis one must choose $m=2$ for $\lambda$ as there are no Clifford operators with $m=1$, $d=3$~\cite{campbell2012magic}. 

The set of operators $U_v$ for $d=3$ is given by:
\begin{equation}\label{Uv3'}
U_v=\sum_{k=0}^{2} \xi^{v_k}\ket{k}\bra{k}.
\end{equation}
where $\xi=e^{2\pi i/9}$. The $v_k$ are given by:
\begin{equation}\label{Uv3b'}
v=(v_0,v_1,v_2)=(0,6z'+2\gamma'+3\epsilon',6z'+\gamma'+6\epsilon'),
\end{equation}
where all operations can be taken modulo $9$. The determinant of $U_v$ for $d=3$ can be computed from this definition:
\begin{equation*}
\det U_v = e^{\frac{2\pi i}{9}\sum_k=0^2 v_k} =  e^{\frac{2\pi i}{3}(z'+\gamma')}
\end{equation*}
showing that $U_v$ is not in $SU(3)$ for $d=3$. 

We can relate the diagonal operators $U_v$ defined by Vala and Howard and the operators $M$ defined by Campbell et. al as follows. Writing:
\begin{equation}
M = \sum_{k=0}^{d-1} \exp(\frac{2\pi i}{d^m}\lambda_k)\ket{k}\bra{k}=\sum_{k=0}^{d-1} \omega^{\lambda_k/d^{m-1}}\ket{k}\bra{k}
\end{equation}
and:
\begin{equation}
U_v = \sum_{k=0}^{d-1} \omega^{v_k} \ket{k}\bra{k}    
\end{equation}
we wish to compare:
\begin{equation}
\frac{\lambda_k}{d^{m-1}}= 
\frac{1}{d}\left[d\binom{k}{3}-k\binom{d}{3}+\binom{d+1}{4}\right]\end{equation}
and
\begin{equation}
v_k = \bar{12}k\{\gamma'+k[6z'+(2k-3)\gamma']\}+k\epsilon'.
\end{equation}

These are both cubic in $k$ so we can find the particular $U_v$ that corresponds to $M$ by equating the coefficients. We begin by setting $k=0$ to find the constant term. We immediately obtain:
\begin{equation}
v_0=0,~~~~~\frac{\lambda_0}{d}=\frac{1}{d}\binom{d+1}{4}    
\end{equation}
We conclude that $U_v$ and $M$ will only be equivalent up to a global phase determined by this convention.

Equating the cubic terms yields $\gamma'=1$. Equating the quadratic terms gives 
\begin{equation}
z'-\frac{\gamma'}{2}=d-1    
\end{equation}
so that $z'=(d-1)/2$. Finally, equating the linear terms gives:
\begin{equation}
\epsilon' = \bar{12}(6d-2d^2-1).    
\end{equation}
We may therefore relate $U_v(z',\gamma',\epsilon')$ and $M$ for arbitrary $d>3$ as follows:
\begin{equation}
M_d=\omega^{\frac{1}{d} \binom{d+1}{4}} U_v((d-1)/2,1,\bar{12}(6d-2d^2-1))  \label{Connd} 
\end{equation}
The first two cases of this equivalence are for $d=5$ amd $d=7$ and, up to a global phase, are as given in equations (70) and (71) of~\cite{howard2012qudit}. 

The case of $d=3$ is distinct ($\bar{12}$ does not exist modulo $3$.) but from the definition of $U_v$ for $d=3$ given in eqn.~\ref{Uv3'} and eqn.~\ref{Uv3b'} we have:
\begin{equation}
M_3=e^{\frac{2\pi i}{9}}U_v(1,1,0) \label{Conn3}
\end{equation}
This is, up to a global phase, as given in eqn. (69) of~\cite{howard2012qudit}.

\subsection{Qudit Magic states}

The gates $M$ also allow us to find eigenstates of $C_M$ as follows. Define the state $\ket{M_k}=M\ket{+_{k}}$, where $\ket{+_k}$ is the eigenstate of $X$ with eigenvalue $\omega^k$. We can calculate:
\begin{equation}
\begin{split}
C_M\ket{M_k}
&\propto MXM^{\dagger}\ket{M_k} \\
&=MXM^{\dagger}M\ket{+_k} \\
&=\omega^{k}M\ket{+_k}  \\
&=\omega^{k}\ket{M_k}
\end{split}
\end{equation}

Given eq.(\ref{UDU2}), Vala and Howard recovered the definition of the magic states of Campbell and showed that these magic states $U_v\ket{+}$ are eigenstates of $C_{\gamma',(1,z')^T}$ with eigenvalue $\omega^{-\epsilon'}$:
\begin{equation}
\begin{split}
& C_{\gamma',(1,z')^T}U_v\ket{+}=\omega^{-\epsilon'}U_v D_{(1|0)}{U_v}^{\dagger}U_v\ket{+}
\\
&=\omega^{-\epsilon'}U_v D_{(1|0)}\ket{+}=\omega^{-\epsilon'}U_v\ket{+}
\end{split} \label{msde}
\end{equation} 

\section{Strong Simulation for qubits.}\label{AppBG}

We review here the strong simulation algorithm given by Bravyi and Gosset in~\cite{bravyi2016improved}.
\begin{figure}[h]
\centering\includegraphics[height=2cm]{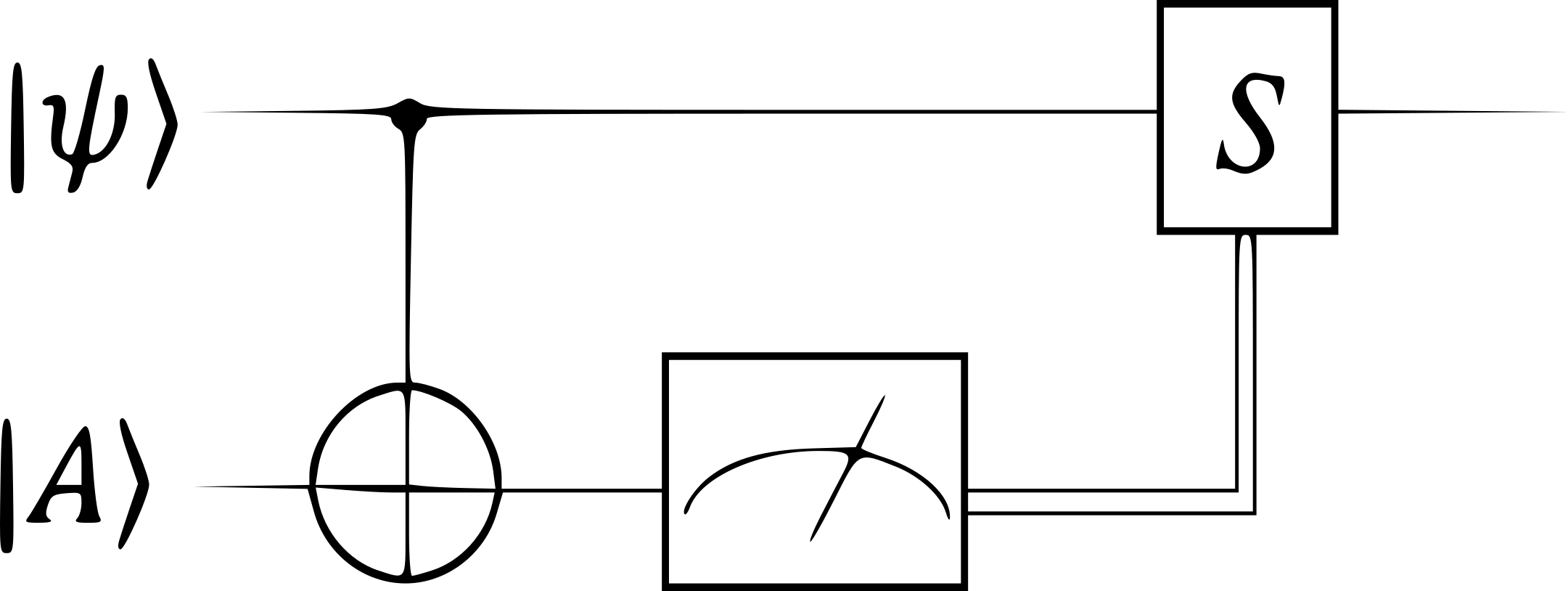}
\caption{Gadget to implement a $T$-gate using an ancilla magic state $\ket{A}$ as defined in~\cite{zhou2000methodology}. Using this gadget, universal quantum computation (UQC) can be achieved using a Clifford circuit with injected magic states.\label{gadgetfig}}
\end{figure}

Let $t$ be the number of $T$ gates in the $n$-qubit quantum circuit we wish to classically simulate. The first step is to replace every $T$ gate in the circuit by Clifford gates and an ancilla input of a magic state $\ket{A}$, defined in~\cite{bravyi2005universal} as:
\begin{equation}\label{AMagic}
\ket{A}=\frac{1}{\sqrt{2}}(\ket{0}+e^{i\pi/4}\ket{1}).    
\end{equation}
This is accomplished using the gadget shown in Figure~\ref{gadgetfig}~\cite{zhou2000methodology}. The number of ancilla qubits is $t$. We consider an initial state $\ket{0^{\otimes n}}$ for the Clifford+$T$ circuit and $\ket{0^{\otimes n}}\otimes\ket{A^{\otimes t}}$ for the gadgetized circuit.

At the end of the computation we will measure $w$ of the $n$ qubits in the logical basis. This measurement with outcome $x$ (where $x$ is a bitstring of length $w$), postselected to the case where all ancilla measurements have result $0$, is represented by a projector $\Pi(x)=\ket{x}\bra{x}\otimes {\bf 1}\otimes \ket{0^t}\bra{0^t}$. The strong simulation algorithm  classically computes the probability of this measurement outcome after acting with a Clifford circuit $V$, which is our original (non-Clifford) circuit with all $T$-gates replaced by the gadget of Figure~\ref{gadgetfig}. Therefore we can express the probability of obtaining output $x$ as:
\begin{equation}
P(x)=2^t\bra{0^n A^t}V^{\dagger}\Pi V\ket{0^n A^t}. \label{Probability}
\end{equation}
The factor of $2^t$ here compensates for the fact that we postselected on the measurement outcomes of the $t$ ancilla qubits.

We define a $t$-qubit projection operator $\Pi_{G}=\bra{0^n}V^{\dagger}\Pi V\ket{0^n}$. This projector maps states onto a stabilizer subspace. Then eq.(\ref{Probability}) becomes
\begin{equation}
\begin{split}
P(x)=2^t\bra{0^n A^t}V^{\dagger}\Pi V\ket{0^n A^t}=2^{-u}\bra{A^t}\Pi_{G}\ket{A^t}. \label{Prob2}
\end{split}
\end{equation}
where $u$ is an integer that depends on the number of qubits we are measuring out of $n$ and the dimension of the stabilizer subspace $\Pi_{G}$ is mapping onto.

If we can expand $\ket{A^t}$ into a sum of stabilizer states, then we can express $P(x)$ as a sum of inner products of $t$-qubit stabilizer states, which can be computed in $O(t^3)$ time~(\cite{aaronson2004improved,garcia2012efficient,bravyi2016trading,bravyi2016improved}). The fewer stabilizer states in the expansion of $\ket{A^t}$, the more efficient the algorithm is. 

{\em Stabilizer rank} is defined as the minimal number of stabilizer states needed to write a pure state as a linear combination of stabilizer states. The value of $\chi(t)$ is trivially upper bounded by $2^t$ because logical basis states are stabilizer states, and $\chi(t)$ is also believed to be lower bounded by an exponential in $t$. For practical purposes we can achieve progress through a series of constructive upper bounds.

In \cite{bravyi2016trading}, Bravyi {\em et al}. found a stabilizer rank upper bound by obtaining $\chi_A(6)\leq7$ for $\ket{A^{6}}$ and dividing the $t$-qubit state into a product of $6$-qubit states. Therefore, $\chi_A(t)$ has a upper bound $7^{t/6}\simeq 2^{0.47t}$. 

If we denote the stabilizer rank for the tensor product of $t$ single-qubit magic states $\ket{A^{t}}$ as $\chi_A(t)$, the cost of classically computing $P(x)$ by taking inner products as described above is $O(t^3{\chi_A(t)}^2)$. 

The quadratic dependence on stabilizer rank can be improved by a  Monte Carlo method, developed by Bravyi and Gosset, to approximate the norm of a tensor product of magic states projected on a stabilizer subspace:
\begin{equation}
|\bra{A^t}\Pi_G\ket{A^t}|=\norm{\Pi_G\ket{A^t}}^2=\norm{\psi}^2
\end{equation}
therefore enabling one to calculate $P(x)$ with cost $O(t^3\chi_A(t))$, linear in stabilizer rank. This concludes our summary of the strong simulation algorithm of Bravyi and Gosset. 

\section{Qudit $T$ gate Gadget}\label{gadget}

\begin{figure}[h]
\includegraphics[height=4cm]{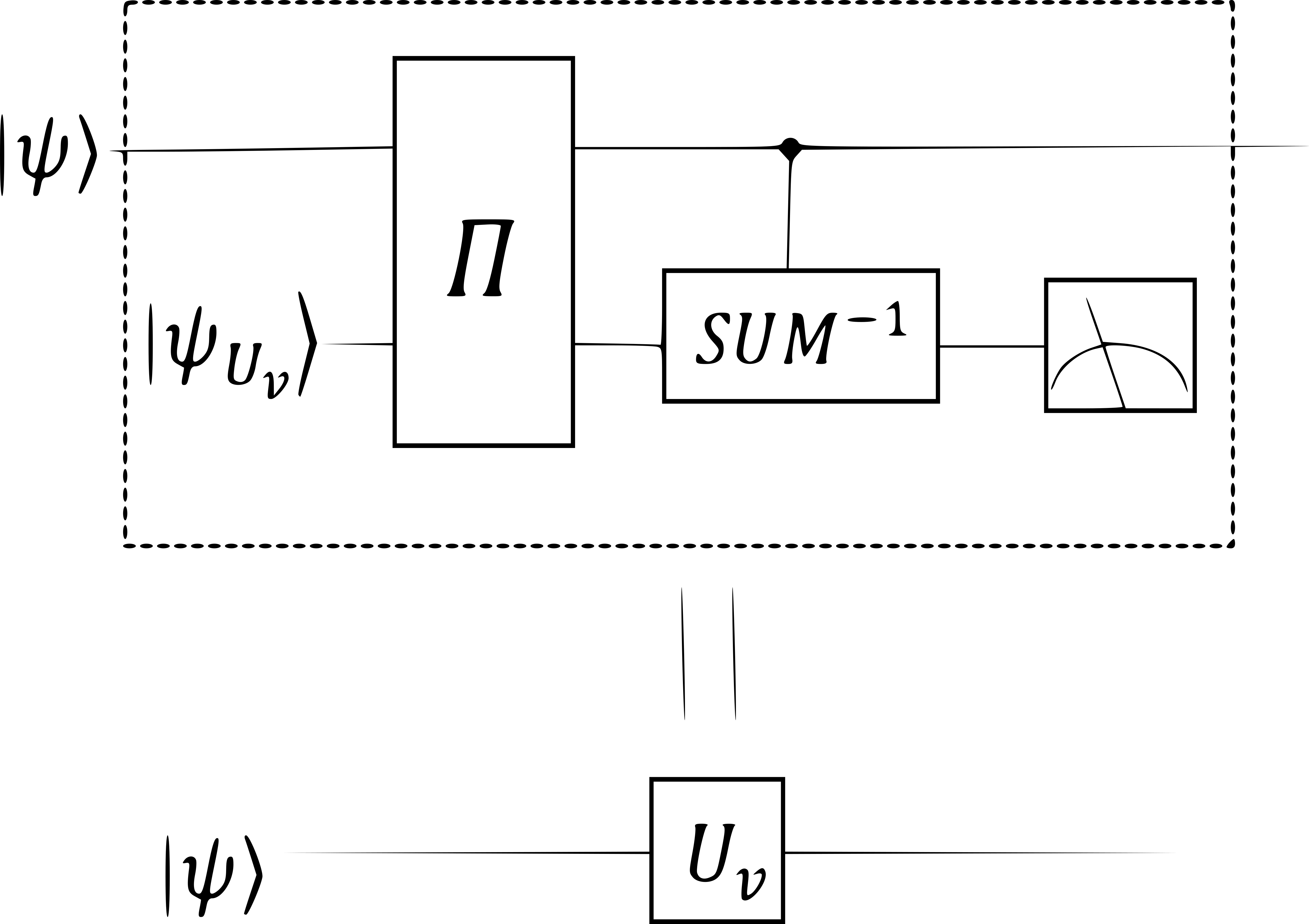}
\caption{Gadget for qudit $U_v$ gate.\label{gadgetfigd}}
\end{figure}

We also require a gadget that substitutes a qudit T-gate by an injected qudit magic state and Clifford gates. The qudit gadget was introduced by Howard and Vala and is shown in Figure~\ref{gadgetfigd}.

Howard and Vala also generalized the qubit T-gate gadget to qudits for their magic state construction~\cite{howard2012qudit}. We reproduce their gadget here in the interest of making the paper self contained. 

In order to project a qudit state onto the eigenstate of operator $P$ with eigenvalue $\omega^{k}$, the projection operator can be written as:

\begin{equation}
\Pi_{(P|k)}=\frac{1}{d}(I+\omega^{-k}P+\omega^{-2k}P^2+...+\omega^{-(d-1)k}P^{d-1})
\end{equation}

By analogy with the qubit case, we need a gadget that allows us to implement qudit $U_v$ gate by injecting magic states. It's straightforward to check that the following performs this task:
\begin{equation}
\text{CSUM}^{-1}\cdot\Pi_{(0,0|1,d-1)[0]}(\ket{\psi}\ket{\psi_{U_v}})=U_v\ket{\psi}\ket{0}
\end{equation}
for a given arbitrary state $\ket{\psi}$, where $\ket{\psi_{U_v}}=U_v\ket{+}$ is the magic state and $\Pi$ is a rank-$p$ projector defined by
\begin{equation}
\Pi_{(0,0|1,d-1)[0]}=\frac{1}{d}[I+Z\otimes Z^{-1}+...+(Z\otimes Z^{-1})^{d-1}]
\end{equation}

This projection is equivalent to measuring the $Z\otimes Z^{-1}$ observable to get eigenvalue 1. If we get eigenvalue $\omega^{k}$, we perform a $X^{-k}$ on the first qudit state to recover it back to the 1-eigenspace. In fact, this gadget works for implementing any diagonal gate $U$ by injecting the state $U\ket{+}$.
 
\section{Magnitude and phase of magic state inner product with orbit representatives of nonorthogonal decompositions\label{appinner}}

Here we compute equation~\ref{inprod}. We begin with $d=3$. In this case we need only tabulate the inner product for three values of $p$:
\begin{equation}
\begin{split}
\bra{+}Z^{-p}\ket{M_3} &=  \bra{+}Z^{-p}M_3\ket{+}\\
&=\frac{1}{d}{\rm Tr} (Z^{-p}M_3)\\
&=\frac{1}{d}\left(e^{2\pi i/9}+e^{-2\pi i p/3}+e^{2\pi i/3(2p-\frac{1}{3})}\right).\\
\end{split}
\end{equation}
Giving:
\begin{equation}
\begin{split}
\bra{+}Z^{0}\ket{M_3}&=\frac{1}{3} \left(1+2 \cos \left(\frac{2 \pi }{9}\right)\right)\\
\bra{+}Z^{-1}\ket{M_3}&=\frac{1}{3} e^{\frac{i \pi }{3}} \left(2 \cos \left(\frac{\pi }{9}\right)-1\right)\\
\bra{+}Z^{-2}\ket{M_3}&=\frac{1}{3} e^{\frac{2i \pi }{3}} \left(1+ 2 \cos \left(\frac{4\pi }{9}\right)\right).\\
\end{split}
\end{equation}
The largest magnitude overlap is obtained for $p=0$.

Now we consider general prime $d>3$. Given the expression for $M_d$ we can write:
\begin{equation}
\begin{split}
\bra{+}Z^{-p}\ket{M_d}= \frac{\omega^{\frac{1}{d}{d+1\choose 4}}}{d}\sum_{j=0}^{d-1}\omega^{\phi(j)}   
\end{split}
\end{equation}
where $\phi(j)$ is a cubic in $j$ given by:
\begin{equation}
\phi(j)={j\choose 3}-\frac{j}{d}{d\choose3}-pj.
\end{equation}
The evaluation of cubic gauss sums is not as straightforward as for quadratic gauss sums. However, we can obtain a closed form for the phase of the sum, up to a sign, by depressing the cubic to remove the quadratic term. In this case this is particularly simple:
\begin{equation}
\begin{split}
\phi'(j)&=\phi(j+1)\\
&={j+1\choose 3}-\frac{j+1}{d}{d\choose3}-pj-p\\
&=\bar 6j\left( j^2-1-(d-1)(d-2)-6p\right)-\frac{1}{d}{d\choose3}-p\\
&=\bar 6j\left( j^2-\psi(d,p)\right)-\frac{1}{d}{d\choose3}-p,
\end{split}
\end{equation}
where:
\begin{equation}
\psi(d,p) = d^2-3d+3 +6p.
\end{equation}
Then:
\begin{equation}
\begin{split}
\bra{+}Z^{-p}\ket{M_d}= \frac{\omega^{\frac{1}{d}{d+1\choose 4} -\frac{1}{d}{d\choose3}-p} }{d}\sum_{j=0}^{d-1}\omega^{\bar 6j\left( j^2-\psi(d,p)\right)}   
\end{split}
\end{equation}

The magnitude of this expression can be determined from the sum, which is real:
\begin{equation}
\begin{split}
S&=\frac{1}{d}\sum_{j=0}^{d-1}\omega^{\bar 6j\left( j^2-\psi(d,p)\right)}\\   &=\frac{1}{d}+\frac{1}{d}\sum_{j=1}^{(d-1)/2}\omega^{\bar 6j\left( j^2-\psi(d,p)\right)}   +\frac{1}{d}\sum_{j=(d+1)/2}^{d-1}\omega^{\bar 6j\left( j^2-\psi(d,p)\right)} \\
&=\frac{1}{d}+\frac{1}{d}\sum_{j=1}^{(d-1)/2}\left(\omega^{\bar 6j\left( j^2-\psi(d,p)\right)}   +\omega^{\bar 6(d-j)\left( (d-j)^2-\psi(d,p)\right)} \right)\\  
&=\frac{1}{d}+\frac{1}{d}\sum_{j=1}^{(d-1)/2}\left(\omega^{\bar 6j\left( j^2-\psi(d,p)\right)}   +\omega^{-\bar 6j\left( j^2-\psi(d,p)\right)} \right)  \\
&=\frac{1}{d}+\frac{2}{d}\sum_{j=1}^{(d-1)/2}\cos\frac{2\pi}{d} \bar 6j\left( j^2-\psi(d,p)\right) 
\end{split}
\end{equation}
While this shows that the sum is real, it does not guarantee that it is positive, and hence the phase of the inner product, up to a sign, is given by:
\begin{equation}
\omega^{\frac{1}{d}{d+1\choose 4} -\frac{1}{d}{d\choose3}-p} = \omega^{\frac{1}{d}{d\choose 4}-p}.
\end{equation}

\section{Canonical forms for Qudit stabilizer states and inner product algorithm}\label{stabilizer}

A qubit stabilizer state can be written as the following canonical form~\cite{nest2008classical,dehaene2003clifford}:
\begin{equation}\label{formqubit}
\ket{\psi}=2^{-m/2}\sum_{x\in \mathcal{A} } (-1)^{q(x)} i^{l(x)}\ket{x}, 
\end{equation} 
where $l(x)$ is a linear form and $q(x)$ takes the quadratic form $q(x)=\sum_{i\neq j} q_{ij}x_{i}x_{j}+c_{i}x_{i}$. $q_{ij}$s, $c_i$s are constants in $Z_2$. $\mathcal{A}$ is an affine space defined as $\mathcal{A}=\left\{Gu+h|u\in{Z_2}^m, h\in{Z_2}^n\right\}$, with $G$ being a $n\times m$ matrix with entries in $Z_2$.

To prove this canonical form holds true for all qubit stabilizer states, one only need to make sure that every state in this form is the eigenstate of a stabilizer operator, as shown in \cite{dehaene2003clifford}. It also suffices to verify that any of the $\{H,P, CNOT\}$ gates preserves the form, only changing the coefficients of $q(x)$, $l(x)$ and affine space $A$. This proof is given in \cite{nest2008classical}.

The normal form was generalized to arbitrary dimensions in~\cite{hostens2005stabilizer}. The stabilizer canonical form for qudits is:
\begin{equation}
\ket{\psi}\propto \sum_{u\in Z_d^k} \omega^{q_d(u)+q_n(u)}\ket{Gu+h}. \label{QuditSForm}  
\end{equation}
where $q_n(u)=\sum_{i\neq j} q_{ij}u_{i}u_{j},~~q_d(u)=\sum_{i=1}^{k} q_i\frac{u_i(u_i-1)}{2}+l_{i}u_{i}, q_{ij}, q_i, l_i\in Z_d$. The state has support in an $k$-dimensional affine space
\begin{equation}
\vec{x}=Gu+h={\rm span}(g^1,...g^k)\oplus h=u_1g^1\oplus u_2g^2...\oplus u_kg^k\oplus h.
\label{affspace}
\end{equation}
$G$ is an $n\times k$ matrix and has each of its columns being $g_1,...,g_k$ with binary entries, while $h$ is a $n\times 1$ vector that has entries in $Z_d$. The division of the phase into two quadratic terms reflects the action of the phase and Hadamard gates, respectively. States of this form were shown to be the $+1$ eigenstate of some Pauli (Weyl-Heisenberg) operator in~\cite{hostens2005stabilizer}. 

This quadratic form on the exponent can also be represented in matrix form:
\begin{equation}
q_d(u)+q_n(u)=2^{-1}u^TQu+Lu. \label{MatrixQuad} 
\end{equation}
where $2^{-1}$ is taken modulo $d$. Here $Q$ is a $k\times k$ matrix with its diagonal terms being $q_i$ and off-diagonal terms being $q_{ij}$ and $L$ is a $1\times n$ matrix where each term corresponds to $l_i-q_i$. 

We will give a new proof that this form is preserved under Clifford operations using the properties of quadratic Gauss sums. We give this proof in order to develop the techniques we will use in the inner product algorithm for qudit stabilizer states.

We consider the single qudit case first. We will prove that the form: 
\begin{equation}\label{quditformG}
\frac{1}{\sqrt{d}}\sum_{j\in Z_d^m}\omega^{f\frac{(j-1)j}{2}+gj}\ket{j+y},
\end{equation}
is preserved under the action of the single-qudit Clifford generators where $f$ and $g$ belong to $Z_d$, $y$ is a shift vector that also belongs to $Z_d$. We are studying single-qudit case here so $m$ is either 0 or 1. When $m=0$, this is simply a computational basis state. 

Acting with diagonal Clifford gates on (\ref{quditformG}) such as $P$ or $Z$ will only change the coefficients $f$ and $g$ in this expression. Similarly, acting with powers of the $X$ gate will only shift $y$, again preserving the quadratic form of the exponents. 

It only remains to check the Hadamard gate:
\begin{equation}
\begin{split}
&H\frac{1}{\sqrt{d}}\sum_{j\in Z_d^m} \omega^{\frac{fj(j-1)}{2}+gj}\ket{j+y} \\
& =\frac{1}{d}\sum_{k} \omega^{yk}\bigg( \sum_{j\in Z_d^m}\omega^{\frac{fj(j-1)}{2}+(k+g)j}\bigg) \ket{k} \label{gausssum}
\end{split}
\end{equation}
If $m=0$, the quantity in the parentheses is simply a phase factor without the sum. Then this form reverts to (\ref{quditformG}) with $f=0~\mod d$. If $m=1$, we recognize the quantity in the parentheses as a Gauss sum. There are again two cases. If $f=0~\mod d$, then we have 
\begin{equation}
\sum_{j\in Z_d}\omega^{(k+g)j}=d\delta_{k+g,0}  
\end{equation}
Then (\ref{gausssum}) reverts to (\ref{quditformG}) as $m=0$ case, i.e., a computational basis state.

If $f\neq 0~\mod d$, to compute this Gauss sum, we first complete the square:
\begin{equation}
\begin{split}
\sum_{j}\omega^{\frac{fj(j-1)}{2}+(k+g)j} 
& = \sum_{j}\omega^{\frac{f}{2}(j^2-j+2(k+g)j\bar{f})}\\ 
& =\omega^{-\bar{2}f(\bar{f}(k+g)-\bar{2})^2}\sum_{j}\omega^{\bar{2}f(j-\bar{2}+(k+g)\bar{f})^2}\\ &=\omega^{-\bar{2}f(\bar{f}(k+g)-\bar{2})^2}\sum_{n}e^{2\pi i\bar{2}fn^2/d}\\ 
\end{split}\label{gausssum1}
\end{equation}
where $\bar{2}$, $\bar{f}$ meaning that $2\bar{2}\equiv 1~{\rm mod}~d$ and $f\bar{f}\equiv 1 ~{\rm mod}~d$.  

The value of this Gauss sum is well known:
\begin{equation}
\sum_{n}e^{2\pi i\bar{2}fn^2/d}=
\begin{cases}
(\frac{\bar{2}f}{d})\sqrt{d}, & d\equiv 1({\rm mod}~4)\\
i(\frac{\bar{2}f}{d})\sqrt{d}, & d\equiv 3({\rm mod}~4),
\end{cases}\label{gausssum2}
\end{equation}
where $(\frac{\bar{2}f}{d})$ is the Legendre symbol.

Hence:
\begin{equation}
\sum_{j}\omega^{\frac{fj(j-1)}{2}+(k+g)j}\propto \omega^{\frac{-\bar{f}k(k-1)}{2}-\bar{2}((2g+1)\bar{f}-1)k}
\label{gausssum3}
\end{equation} 
The new coefficients $-\bar{f}$ and $-\bar{2}((2g+1)\bar{f}-1)$ here are still in $Z_d$. This means that the general form $\sum_{k}\omega^{f\frac{(k-1)k}{2}+gk}\ket{k}$ of single-qudit stabilizer states is preserved under the action of any Clifford operations.  

For multi-qudit states, we have the same affine space property as the qubit case except that the additions are modulo $d$. Before we give the proof, we need to show that quadratic form given in terms of the basis vectors of the affine space $\vec{u}$ and the qudit vectors itself $\vec{x}$ are equivalent. Changing the arguments only changes the coefficients of the quadratic form. Given eq.(\ref{MatrixQuad}), we further assume the quadratic and linear matrices in terms of $x$ being $\tilde{Q}$ and $\tilde{L}$:
\begin{equation}
\begin{split}
\omega^{x^T\tilde{Q}x+\tilde{L}x} &=\omega^{(u^TG^T+h^T)\tilde{Q}(Gu+h)+\tilde{L}(Gu+h)}\\    
&\propto \omega^{u^TG^T\tilde{Q}Gu+(2h^T\tilde{Q}G+\tilde{L}G)u}
\end{split}
\end{equation}
From this equation, we can see the relationship between $Q$, $L$ and $\tilde{Q}$, $\tilde{L}$: $Q=G^T\tilde{Q}G$, and $L=2h^T\tilde{Q}G+\tilde{L}G$.

Now we use Van den Nest's method~\cite{nest2008classical} to prove that the canonical form (\ref{QuditSForm}) is preserved under the action of $CSUM$, $P$ and $H$. The $CSUM_{i\rightarrow j}$ gate shifts the affine space by mapping $\ket{a}\ket{b}$ to $\ket{a}\ket{a\oplus b}$, without changing the phases. As in the qubit case, we only need to add the $i$th column of the matrix $G$ to the $j$th column:
\begin{equation}
\begin{split}
CSUM\sum_{u\in Z_d^m} \omega^{q_d(u)+q_n(u)}\ket{Gu+h}
& \\=\sum_{u\in Z_d^m} \omega^{q_d(u)+q_n(u)}\ket{G'u+h}.
\end{split}
\end{equation}
$G'$ differs from $G$ by $g_j\rightarrow g_i\oplus g_j$.

Acting with $P$ on qudit $i$ results in the state:
\begin{equation}
P_i\ket{\psi}\propto\sum_{x\in A}\omega^{q_n(x)+q_d(x)}\omega^{\frac{x_i(x_i-1)}{2}}\ket{x}   
\end{equation}
which again leaves the canonical form unchanged. 

The Hadamard gate requires some work. Without loss of generality, we assume that $H$ acts on the first qudit:
\begin{equation}
H_1\ket{\psi}\propto \sum_{v=0}^{d-1}\sum_{u} \omega^{q_n(u)+q_d(u)+v(\bar{g_1}u+h_1)}\ket{v,\bar{G}u+\bar{h}} \label{HadamardonQudit}
\end{equation}
where $\tilde{g_1}^T$ is the first row of $G$ and $\bar{G}$ is the rest of it. If $\bar{G}$ is still full rank after taking out $\tilde{g_1}^T$, we obtain the new $G'$ to be:
\begin{equation}
\begin{pmatrix}
1 & \vec{0}^T \\ \vec{0} & \bar{G}
\end{pmatrix}
\end{equation}
Therefore we have $m+1$ basis vectors now, and $v$ becomes the new $u_1$. The term $v(\bar{g_1}u+t_1)$ in the phase can be absorbed in the quadratic form $q_n(u)$. So this is of the canonical form (\ref{QuditSForm}).

If $\bar{G}$ is rank $m-1$ after taking out $\tilde{g_1}^T$, then the columns of $\bar{G}$ are not linearly independent. In this case one of the $u_i$s is redundant and we want it to be summed out in order to get back to the canonical form. Without loss of generality, let's assume that $u_1=\sum_{i=2}^{m}r_i\bar{g_i}$, therefore $\bar{G}u+\bar{h}=\sum_{i=2}^{m}(u_i+r_i)\bar{g_i}+\bar{h}$. If we denote $u_i'\equiv u_i+r_i$ for $i=2$ to $m$~($\bar{u}$) and $u_1'\equiv v$, $q_n(\bar{u})$ and $q_d(\bar{u})$ can be written in terms of $\bar{u'}$ with different coefficients from $q_n$ and $q_d$, say $q_n'(\bar{u'})$ and $q_d'(\bar{u'})$, together with some constant factor which can be neglected. Then eq.(\ref{HadamardonQudit}) becomes:
\begin{equation}
\begin{split}
& H_1\ket{\psi}\propto \sum_{v=0}^{d-1}\sum_{u} \omega^{q_n(u)+q_d(u)+v(\tilde{g_1}^Tu+h_1)}\ket{v,\bar{G}u+\bar{h}}\\ &\propto \sum_{v=0}^{d-1}\sum_{u_2',...u_m'}\sum_{u_1} \omega^{q_n(u)+q_d(u)+v(\tilde{g_1}^Tu+h_1)}\ket{v,\sum_{i=2}^{m}u_i'\bar{g_i}+\bar{h}} \\ 
&=\sum_{u_1',u_2',...u_m'}\omega^{q_n'(\bar{u}')+q_d'(\bar{u}')+u_1'(\sum_{i=2}^{m}g_{1i}(u_i'-r_i)+\bar{h_1})}\\ 
&\bigg(\sum_{u_1}\omega^{q_n(u_1)+q_d(u_1)+vg_{11}u_1}\bigg)\ket{v,\sum_{i=2}^{m}u_i'\bar{g_i}+\bar{h}}
\end{split}
\end{equation}
Here the parenthesis contains the Gauss sum we computed earlier. Then we can drop the prime for the $u$s and absorb the result of the Gauss sum and $u_1'(\sum_{i=2}^{m}g_{1i}(u_i'-r_i)+\bar{h_1})$ into the $q_n'$ and $q_d'$ functions. Finally we arrive at the same form but with different coefficients. Hence, the canonical form is preserved under the action of all Clifford gates.

We now use this canonical form and the Gauss sum techniques to provide an $O(n^3)$ algorithm for the computation of the inner products of two qudit stabilizer states.

\subsection{The inner product of two qudit stabilizer states}\label{inner}
The inner product between two qubit stabilizer states can be computed efficiently in $O(n^3)$~(\cite{aaronson2004improved,garcia2012efficient,bravyi2016trading,bravyi2016improved}). However, a corresponding algorithm for qudits has not yet been given, although most aspects of the theory of stabilizer states have been generalized ~\cite{gottesman1999fault,hostens2005stabilizer}. We will now describe a $O(n^3)$ algorithm that computes the inner product of two qudit stabilizer states based on the Gauss sum  techniques we discussed in the previous section. 

As discussed above, the quadratic form in terms of the basis vector of the affine space $\vec{u}$ and the qudit vector itself $\vec{x}$ are equivalent. Therefore eq. (\ref{QuditSForm}) is equivalent to the following:
\begin{equation}
\ket{\psi}\propto \sum_{x\in A}\omega^{\tilde{q_n}(x)+\tilde{q_d}(x)}\ket{x}
\label{QuditForm2}
\end{equation}
where $A$ is the affine space defined by $Gu+h$ in eq.(\ref{affspace}).

Assume we have two qudit stabilizer states $\ket{\psi_1}$ and $\ket{\psi_2}$, which take the above form (\ref{QuditForm2}) with subindices 1 and 2:
\begin{equation}
\begin{split}
\bra{\psi_2}\ket{\psi_1}
&= d^{-(k_1+k_2)/2}\sum_{x_1\in A_1}\sum_{x_2\in A_2}\omega^{\tilde{q_1}(x_1)-\tilde{q_2}(x_2)}\braket{x_2}{x_1} \\
&= d^{-(k_1+k_2)/2}\sum_{x\in A_1\cap A_2}\omega^{\tilde{q_1}(x)-\tilde{q_2}(x)} \\
&= d^{-(k_1+k_2)/2}\sum_{x\in A_1\cap A_2}\omega^{\tilde{q}(x)}\\
&= d^{-(k_1+k_2)/2}\sum_{u\in F_d^k}\omega^{q(u)} \label{InnerprodA_1A_2}
\end{split}
\end{equation}
where $\tilde{q_1}=\tilde{q_{1d}}+\tilde{q_{1n}}$, $\tilde{q_2}=\tilde{q_{2d}}+\tilde{q_{2n}}$, $\tilde{q}=\tilde{q_1}-\tilde{q_2}$, $k$ is the dimension of $A_1\cap A_2$ and $q$ is the quadratic form in the new basis of $A_1\cap A_2$. The new basis of the affine space $A_1\cap A_2$, as well as the new quadratic form associated with it, can be calculated with the same method used by Bravyi and Gosset in Appendix $B$, $C$ for qubits~\cite{bravyi2016improved}, with cost $O(n^3)$. 

What remains in eq.(\ref{InnerprodA_1A_2}) is a Gauss sum, which we again rewrite in the following form:
\begin{equation}
\sum_{u\in F_d^k}\omega^{u^{T}Qu+Lu}
\end{equation}
where the exponent is given by eq. (\ref{MatrixQuad}). We can diagonalize $Q$ and factor this sum into a product of $k$ Gauss sums over $F_d$. We obtain a transformation matrix $P$ that gives:
\begin{equation}
P^{T}QP=\Lambda, \label{Diagonalization'}
\end{equation}
where $\Lambda$ is the diagonal matrix with entries $(\lambda_1,...,\lambda_k)$. 

Then if we further define $u=Pu'$, we obtain
\begin{equation}
\begin{split}
\sum_{u\in F_d^k}\omega^{u^{T}Q u+Lu}
&=\sum_{u'\in F_d^k}\omega^{u'^{T}P^{T}QPu'+LPu'}\\
&=\sum_{u'\in F_d^k}\omega^{u'^{T}\Lambda u'+LPu'}\\
&=\prod_{i=1}^{k}\sum_{u_i\in F_d}\omega^{\lambda_i u_i'^2+l_i'u_i'},
\end{split}
\end{equation}
where $l_i'=\sum_j p_{ji}l_j$. This is a product of $k$ Gauss sums, as given in eq.~(\ref{gausssum1}, \ref{gausssum2}, \ref{gausssum3}). 

Each Gauss sum only takes $O(1)$ time, so the product of $k$ of them takes time $O(k)$. The scaling of this algorithm is determined by the complexity of Gaussian elimination, $O(k^3)$ because $Q$ has rank $k$. Therefore, together with the first step to obtain $A_1\cap A_2$, the algorithm takes $O(n^3)$ time overall in the worst case.

\section{Evaluation of $Z(\mathcal{L})$\label{Zeval}}

The quantity $Z)_(\mathcal{L})$ is given by eq.~(\ref{Z(L)}):
\begin{equation}
\begin{split}
Z(\mathcal{L})&=\sum_{x\in \mathcal{L}}\bra{\tilde{0}^t}C_{\vec{x}}\ket{\tilde{0}^t} \\  
&=\sum_{x\in \mathcal{L}}\prod_{l=1}^t\bra{\tilde{0}}C^{x_l}\ket{\tilde{0}}\\  
&=\sum_{x\in \mathcal{L}}\prod_{l=1}^t\bra{\tilde{0}}\ket{\tilde x_l}  
\end{split}
\end{equation}
We can see that this quantity is a function of the values $\braket{\tilde{0}}{\tilde{1}}$,...,$\braket{\tilde{0}}{\tilde{d-1}}$. We label the phase of $\braket{\tilde{0}}{\tilde{j}}$ by $\beta_j$ for all $j$, where $\beta_0=1$. Using eqn.~(\ref{noboverlap})  $Z(\mathcal{L})$ can be rewritten in the following form:
\begin{equation}
\begin{split}
Z(\mathcal{L})&=\sum_{x\in \mathcal{L}}\prod_{l=1}^t\bra{\tilde{0}}\ket{\tilde x_l} \\
&=\sum_{x\in \mathcal{L}}\prod_{l=1}^t \beta_{x_l}\sqrt{\frac{1+(d-1)\delta_{0,x_l}}{d}} \\
&=\sum_{x\in \mathcal{L}} \frac{\Pi_{l=1}^{t} \beta_{x_l}}{d^{(t-\abs{x})/2}}\\ 
&=\sum_{x\in \mathcal{L}} \frac{\Pi_{j=1}^{d-1} \beta_{j}^{\abs{x}_j}}{d^{(t-\abs{x})/2}} \label{explicitZ}
\end{split}
\end{equation}
where $\abs{x}$ is the Hamming weight of codeword $x$ in code $\mathcal{L}$, i.e. the number of nonzero elements in the codeword. $\abs{x}_j$ means the number of digits in string $x$ that equals to $j$. If we regard $\mathcal{L}$ as a linear code, then the qubit case $Z(\mathcal{L})$ is exactly the weight enumerator of the code. In the qudit case, $Z(\mathcal{L})$ depends on the Hamming weight as well as the $\beta_j$s. Now let's calculate an explicit expression for the $\beta_j$s.

For the $d=3$ case, we specifically obtain $\beta_1=e^{\pi i/18}$ and $\beta_2=e^{-\pi i/18}$. For $d>3$ case, we assume our initial stabilizer state $\ket{\tilde{0}}=Z^p\ket{+}$. And
\begin{equation}
\beta_j=\sqrt{d}\braket{\tilde{0}}{\tilde{j}}=\sqrt{d}\bra{\tilde{0}}C^j\ket{\tilde{0}}    
\end{equation}
where the $C$ for Campbell's choice of $\ket{M_d}$ is simply $\omega^{-\bar{3}}XP$ according to eq.(\ref{C}) and subsection C of section \ref{decomposition}. We can calculate $(XP)^j$ as
\begin{equation}
\begin{split}
(XP)^j&=\sum_k \omega^{\sum_{l=0}^{j-1}\binom{k+l}{2}}\ket{k+j}\bra{k}\\
&=\omega^{\bar 6(j^3-3j^2+2j)}\sum_k \omega^{\bar 2(jk^2+(j^2-2j)k)}\ket{k+j}\bra{k}. 
\end{split}
\end{equation}
Therefore we can rewrite $C^j$ as
\begin{equation}
\begin{split}
C^j&=\omega^{-\bar{3}j}(XP)^j\\
&=\omega^{\bar 6(j^3-3j^2)}\sum_k \omega^{\bar 2(jk^2+(j^2-2j)k)}\ket{k+j}\bra{k}. \end{split}
\end{equation}
Then we can calculate $\beta_j$ as
\begin{equation}
\begin{split}
\beta_j&=\sqrt{d}\bra{+}Z^{-a}C^jZ^a\ket{+} \\
&=\frac{\omega^{\bar 6(j^3-3j^2)}}{\sqrt{d}}\bra{k''}\sum_{k''}\omega^{-pk''}\sum_k \omega^{\bar 2(jk^2+(j^2-2j)k)}\ket{k+j}\\
&\bra{k}\sum_{k'}\omega^{pk'}\ket{k'}\\
&=\frac{\omega^{\bar 6(j^3-3j^2-6pj)}}{\sqrt{d}}\sum_k \omega^{\bar 2(jk^2+(j^2-2j)k)}
\end{split}
\end{equation}
This is a quadratic Gauss sum times a phase. Using eq.(\ref{gausssum1}) and (\ref{gausssum2}) for $f=j$ and $k+g=\bar{2}(j^2-j)$, we obtain:
\begin{equation}
\sum_k \omega^{\bar 2(jk^2+(j^2-2j)k)}= \begin{cases}
\omega^{-\bar{2}^3j(j-2)^2}(\frac{2j}{d})\\
\omega^{-\bar{2}^3j(j-2)^2}i(\frac{2j}{d}).
\end{cases}   
\end{equation}
The final expression of $\beta_j$ in terms of $p$ is:
\begin{equation}
\begin{split}
\beta_j &=
\begin{cases}
\omega^{\bar{6}j^3-\bar{2}j^2-pj}\omega^{-\bar{2}^3j(j-2)^2}(\frac{2j}{d})\\
\omega^{\bar{6}j^3-\bar{2}j^2-pj}\omega^{-\bar{2}^3j(j-2)^2}i(\frac{2j}{d})\\
\end{cases}  \\
&=
\begin{cases}
\omega^{(\bar{6}-\bar{2}^3)j^3-(p+\bar{2})j}(\frac{2j}{d}),& d\equiv 1({\rm mod}~4)\\
\omega^{(\bar{6}-\bar{2}^3)j^3-(p+\bar{2})j}i(\frac{2j}{d}),& d\equiv 3({\rm mod}~4)\\
\end{cases}
\end{split}
\end{equation}
where again $(\frac{\bar{2}j}{d})$ is the Legendre symbol. 

\end{document}